\documentclass[11pt]{article}
\usepackage{graphicx}
\usepackage{amssymb}
\usepackage[nomove]{overcite}
\usepackage{amsmath}
\usepackage{amssymb}
\usepackage[small]{caption2}
\usepackage{xspace}
\begin{document}
%
\title{\sffamily \bfseries Adhesion between a viscoelastic material\\ and a solid surface}
\author{F.\ Saulnier$^1$\footnote{florent.saulnier@college-de-france.fr}\ ,
T.\ Ondar\c{c}uhu$^{2}$\footnote{ondar@cemes.fr}\ , A.\ Aradian$^{1,3}$\footnote{A.Aradian@ed.ac.uk}
\ and E.\ Rapha\"{e}l$^1$\footnote{elie.raphael@college-de-france.fr} \\
$^1${\footnotesize Laboratoire de Physique de
la Mati\`ere Condens\'ee UMR CNRS  7125}\\
{\footnotesize and F\'ed\'eration de
Recherche Mati\`ere et Syst\`emes Complexes FR CNRS 2438, }\\
{\footnotesize Coll\`ege de France,
11, place Marcelin Berthelot, 75231 Paris Cedex 05, France}\\
$^2$ {\footnotesize Centre d'Elaboration des Mat\'eriaux et
d'Etudes Structurales (CEMES), U.P.R. CNRS 8011,}\\ {\footnotesize
29, rue Jeanne Marvig, 31055 Toulouse Cedex 4, France.}\\
$^3$ {\footnotesize University of Edinburgh, School of
Physics,King's Buildings JCMB,}\\ {\footnotesize Edinburgh EH9
3JZ, United Kingdom.}}

\maketitle
\begin{abstract}
In this paper, we present a qualitative analysis of the
dissipative processes during the failure of the interface between
a viscoelastic polymer, characterized by a weak adhesion, and a
solid surface. We reassess the "viscoelastic trumpet" model [P.-G.
de Gennes, C. R. Acad. Sci. Paris, {\bf 307}, 1949 (1988)], to
express the viscous energy dissipated in the bulk as a function of
the rheological moduli of the material, involving the local
frequencies of sollicitation during crack propagation. We deduce
from this integral expression the dhesion energy for different
kind of materials: (i) we show that, for a crosslinked polymer,
the dissipation had been underestimated at low velocities. Indeed,
the interface toughness $G(V)$  starts from a relatively low
value, $G_0$, daue to local processes near the fracture tip, and
rises up to a maximum of order $G_0 (\mu_{\infty}/\mu_0)$ (where
$\mu_0$ and $\mu_{\infty}$ stand for the elastic modulus of the
material, respectively at low and high strain frequencies). This
enhancement of fracture energy is due to far-field viscous
dissipation in the bulk material, and begins for peel-rates $V$
much lower than previously thought. (ii) For a polymer melt, the
adhesion energy is predicted to scale as $1/V$. In the second part
of this paper, we compare some of these latest theoretical
predictions with experimental results about the viscoelastic
adhesion between a polydimethylsiloxane polymer melt and a glass
surface. In particular, the expected dependence of the fracture
energy versus separation rate is confirmed by the experimental
data, and the observed changes in the concavity of the crack
profile are in good agreement with our simple model. More
generally, beyond the qulitative and simple picture used for our
approach, we expect our theoretical treatment to apply for
relatively {\it{weak viscoelastic adhesives}}, for which the
crack-tip dissipative term $G_0$ is weakly dependent on the
fracture velocity.
\end{abstract}

\section{Introduction}

Understanding how the interface between a polymer and another
material fails is important for many industrial applications and
has therefore been the subject of many studies in the last 30
years \cite{hugh_brown, jones_richards, creton_kramer_brown_hui}.
A quantity of central interest is the \textit{interface toughness}
(also called the adhesion energy), $G$, which is the energy per
unit area needed to make a crack separating the two materials
travel along the interface. If the polymer is above its glass
transition temperature, this energy is dissipated - as the crack
advances - by both local processes (occurring near the crack tip)
and viscoelastic losses (taking place over macroscopic volumes)
\cite{gent}. Some years ago, Gent and Schultz \cite{gent_shultz}
and Andrews and Kinloch \cite{andrews_kinloch} showed that, for
elastomeric adhesives, the variations of the interface toughness,
G, with the crack velocity, $V$, can be written as:

\begin{equation}
G(V) = G_{0}(1 + \varphi(a_{T} V)), \label{eq Gent}
\end{equation}

where $G_{0}$ is the limiting value of the fracture energy at zero
rate of crack growth, and represents local processes. According to
Eq.(\ref{eq Gent}), the contribution of the bulk viscoelastic
losses, $G_{v}(V) = G(V) - G_{0}$, is given by $G_{v}(V) = G_{0}
\varphi(a_{T} V)$ (where the temperature-shift factor $a_{T}$ is
given by the Williams-Landel-Ferry equation \cite{wlf}) and is
therefore itself proportional to the local contribution $G_{0}$.
This remarkable fact was explained by de Gennes at the level of
scaling laws \cite{PGG1,PGG2}, and further developed more
rigorously by Hui, Xu and Kramer \cite{hui_xu_kramer} (for other
related studies, see \cite{christensen_and_wu, bowen_knauss,
christensen, schapery, freund_hutchinson, barber}). In the first
part of this paper, we reconsider de Gennes' model in the case of
the interface between a poorly crosslinked elastomer (or a polymer
melt) and a solid surface. We show, in particular, that the
far-field viscoelastic contributions to the interface toughness
play a significant role at separation rates much lower than
previously thought. We also reconsider the profile of the crack
\cite{PGG1}, confirming some predictions of the earlier approach
of Greenwood and Johnson \cite{greenwood}. In the second part of
the paper we present experimental results for the adhesion between
a polymer melt and a glass surface (for earlier work on "tack",
see, \emph{e.g.}, refs. \cite{PGG_pegosite, gent_kim, creton_leibler,
gay_leibler}). These results for the fracture energy and
the crack profile, which extend earlier work by Ondar\c{c}uhu
\cite{Ondarcuhu}, are then compared with the theoretical
predictions of section \ref{theoretical}.

\section{Theoretical approach}
\label{theoretical}
\subsection{Viscoelastic features of the polymer material}

Many polymers are characterized by a viscoelastic behavior,
exhibiting liquid-like or solid-like responses to mechanical
sollicitations, depending on the frequency range of sollicitation.
For the sake of simplicity, as in Refs.\cite{PGG1,hui_xu_kramer},
we assume that the polymer material is characterized by a single
relaxation time\cite{relaxation_time}, $\tau$ , and that its
rheological behavior can be described by a complex modulus
$\underline{\mu}(\omega)=\mu'(\omega)+i \mu''(\omega)$ (we denote
complex quantities with underlined variables), given by:

\begin{equation}
\underline{\mu}(\omega) = \mu_{0}+ (\mu_{\infty}-\mu_0) \frac{i
\omega \tau}{1+ i \omega \tau}. \label{rheolaw}
\end{equation}

This expression \ref{rheolaw} can depict both mechanical responses
of crosslinked and uncrosslinked polymers, by taking in this last
case $\mu_0 = 0$ (\emph{i.e.} in the absence of low-frequency
elastic modulus). In the following, we will discuss both cases and
will use a fundamental parameter, denoted $\lambda$,
characterizing the ratio between high and low-frequency elastic
moduli. For a poorly cross-linked elastomer, some chains are tied
by one end only (some others might even be free): in this case,
the low-frequency modulus $\mu_0$ (related to the network) is
small, while the high-frequency modulus $\mu_{\infty}$ (which
contains the effects of the entangled free chains and of the
dangling ends) is high. We will thus suppose that this ratio of
modulus is high, as it can be typically achieved \cite{gent_lai}:

\begin{equation}
\lambda \equiv \frac{\mu_{\infty}}{\mu_0} \sim 100. \label{factor
lambda}
\end{equation}
Comparing the deformation rate with the relaxation time of the
material, $\tau$, {\it three} regimes can be distinguished, as
summarized in Fig.1:
\begin{figure}
\centering\resizebox{0.65 \textwidth}{!}{%
 \includegraphics*[2cm,8.7cm][19.6cm,14cm]{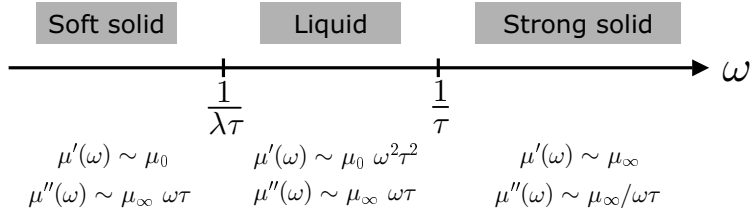} }
  \caption{Viscoelastic features of a cross linked rubber: $\tau$
  is the relaxation time, $\mu_0$ is the elastic modulus at low
  frequencies $\omega$, $\mu_{\infty}$ is the elastic modulus at high
  frequencies, and $\eta$ is the viscosity of the liquid in the
  intermediate frequency range.}
  \label{rheo}
\end{figure}

(i) At very low $\omega$, \emph{i.e.} $\omega<1/(\lambda \tau)$,
we have $\underline{\mu} \approx \mu_0$. The complex modulus is
thus essentially real (\emph{i.e.} $\mu' \gg \mu''$): the elastic
component dominates, and the material can be considered a {\it
soft solid}.

(ii) For $1/(\lambda \tau)<\omega<1/\tau$, we have:
\begin{equation}
\underline{\mu}(\omega) \approx (\mu_{\infty}-\mu_0) i \omega \tau
= i \omega \eta. \label{viscous}
\end{equation}
In this frequency range, where $\mu' \ll \mu''$, the rheology is
mainly viscous-type: the material is a {\it liquid} with a
viscosity $\eta = (\mu_{\infty}-\mu_0) \tau \sim \mu_{\infty}
\tau$.

(iii) At high frequencies, \emph{i.e.} $\omega > 1/\tau$, we find
back a regime where $\mu' \gg \mu''$, and thus recover a {\it
strong solid} with an elastic modulus $\mu \approx \mu_{\infty}$.

During crack propagation, the strain rate imposed to the material
is high near the fracture tip, and lowers as the distance $x$ to
the head increases (far from the tip, the material had more time
to relax the stresses). Following Ref.\cite{PGG2}, we thus relate
the distance $x$ to the tip with the frequency $\omega$ by a
simple scaling law of the form:
\begin{equation}
\omega \cong \frac{2 \pi V}{x}. \label{relation omega x}
\end{equation}

As a consequence, we can distinguish three spatial regions in the
bulk of the moving rubber, corresponding to the three regimes of
frequencies defined above. In Fig.2  we present a simple
view of the fracture profile when the crack propagates in a
crosslinked polymer at speed $V$. Note that this graphic
representation corresponds to a velocity chosen in the range
$l/\tau <V< L/\lambda \tau$, where $L$ is the crack length
\cite{length_L}.
\begin{figure}
\centering\resizebox{0.65 \textwidth}{!}{%
  \includegraphics*[2cm,10.3cm][19cm,22.8cm]{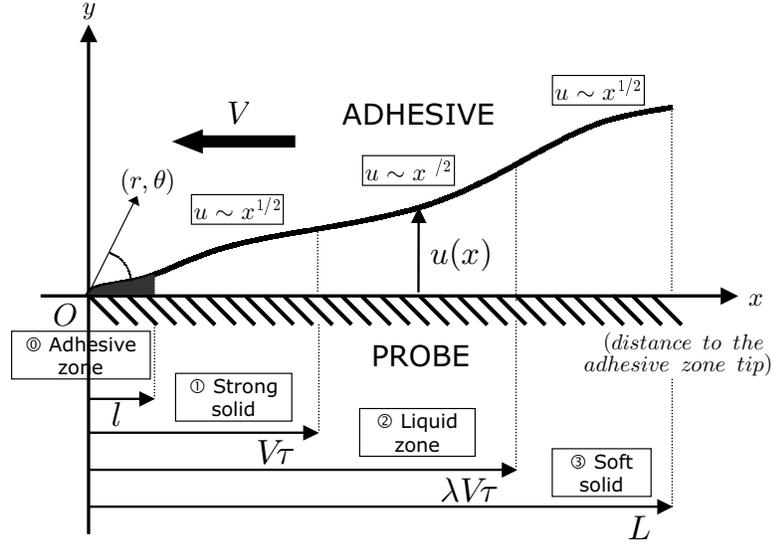} }
  \caption{The "viscoelastic trumpet" model for a crosslinked elastomer characterized
  by a single relaxation time $\tau$.
  By a simple relation between the distance to the fracture tip ($x$) and the
  frequency of deformation ($\omega$), three regions with different viscoelastic
  properties can be distinguished. The
  adhesive zone is a region of length $l$ where high stresses apply and give
  rise to irreversible dissipative processes. The moving frame
  $(x,y)$ is centered at the adhesive zone tip, and $(r,\theta)$
  is the associated polar coordinate system.}
  \label{model}
\end{figure}

Directly ahead the crack tip (located at $x=l$) is an adhesive
zone of length $l$ (assumed to be smaller than $1000 \AA$), where
local dissipative processes take place \cite{local}, and lead to
the $G_0$ term of order 1 to 10 J$\cdot$m$^{-2}$. In the
following, all relevant dimensions of the problem ({\it e.g.} the
crack length or the specimen dimensions in a fracture test) will
be assumed to be sufficiently large compared with the adhesive
zone size. For simplicity reasons, we will also suppose this
length $l$ constant and independent of separation rate $V$


Close to the fracture tip, the small spatial scales correspond to
high rates of deformation: this region (number 1 in
Fig.2 : $x < V \tau$) is a strong solid of elastic
modulus $\mu_{\infty}$. At intermediate distances (region 2:
$V\tau <x < \lambda V \tau$), the behavior is viscous-type: the
polymer can be viewed a liquid of viscosity $\eta$. Far away the
fracture head (region 3: $x>\lambda V \tau$), the material is a
soft solid of modulus $\mu_0$.

\subsection{Integral expression for the fracture energy}

A few years ago \cite{PGG2}, de Gennes proposed a general relation
between the energy $G_v(V)$ viscously dissipated during crack
propagation at separation rate $V$, and the real
($\mu'(\omega)=\mbox{Re}[\underline{\mu}(\omega)]$) and imaginary
($\mu''(\omega)=\mbox{Im}[\underline{\mu}(\omega)]$) parts of the
complex modulus $\underline{\mu}(\omega)$.

\subsubsection{Viscous dissipation $G_v(V)$ in the polymer bulk}

Let us begin with a general calculation of the energy dissipated
in a viscoelastic material, within the framework of linear
viscoelasticity and using the more convenient complex
representation of the oscillatory motion. When the material is
submitted to an oscillatory stress at frequency $\omega$,
characterized by its complex form $\underline{\sigma}=\sigma_0
\exp{i \omega t}$, the material response, given by the complex
strain, $\underline{\gamma}$, is of the form:
\begin{equation}
\underline{\sigma} = \underline{\mu}(\omega) \underline{\gamma}.
\label{sigma mu gamma}
\end{equation}

The energy dissipated per unit of time, and per unit of volume, is
given by $\sigma \dot{\gamma}$, where $\dot{\gamma}$ denotes the
strain-rate. One can show that the time-averaged of this quantity,
$\langle \sigma \dot{\gamma} \rangle $, is simply given by:
\begin{equation}
\langle \sigma \dot{\gamma} \rangle = \mbox{Re} \left[
\frac{\underline{\sigma} \cdot \underline{\dot{\gamma}}^*}{2}
\right]. \label{lien dissipation reelle et complexe}
\end{equation}

Here, $\underline{\dot{\gamma}^*}$ is the complex conjugate of the
complex strain-rate $ \underline{\dot{\gamma}}=i \omega
\underline{\gamma}$.

Using Eq.(\ref{sigma mu gamma}), we find that the viscous energy
dissipated by the system depends on \emph{both} loss ($\mu''$) and
storage ($\mu'$) modulus:

\begin{equation}
\mbox{Re} \left[ \frac{\underline{\sigma} \cdot
\underline{\dot{\gamma}}^*}{2} \right]= \frac{\sigma_0^2}{2} \cdot
\frac{\omega \, \mu''(\omega)}{\mu'(\omega)^2+\mu''(\omega)^2}.
\label{reel sigma gamma point}
\end{equation}

Let us now turn back to the trumpet model. The viscous dissipation
$T \dot{S}$ (per unit length of the fracture line) is:
\begin{equation}
T \dot{S} = \int \int \langle \sigma \dot{\gamma} \rangle \,
\mbox{d}x \mbox{d}y = \int \int \mbox{Re} \left[
\frac{\underline{\sigma} \cdot \underline{\dot{\gamma}}^*}{2}
\right]\, \mbox{d}x \mbox{d}y.  \label{tspoint general}
\end{equation}

In our model, we assume that the stress amplitude is given by
$\sigma_0= K_I/\sqrt{r}$ (cf. part \ref{tromp prof} for
precisions). We can also relate the distance to crack tip $r$ to
local frequencies of excitation $\omega$ by the scaling relation
(\ref{relation omega x}): $\omega(r)=V/r$. The integrand inside
the right-hand side of Eq.(\ref{tspoint general}) thus depends
only on $r$, and we can replace the integral over $x$ and $y$ (on
the half-space $y\geqslant 0$) by an integral over $r$ (from $r=l$
to $r \sim L$), omitting numerical coefficient due to the
integration over $\theta$ (from 0 to $\pi$):

\begin{equation}
T \dot{S} \cong \int \left. \sigma_0^2 \frac{\omega \,
\mu''(\omega)}{\mu'(\omega)^2+\mu''(\omega)^2} \right|
_{\omega=V/r} r \mbox{d}r. \label{tspoint sinus}
\end{equation}

Using relation (\ref{relation omega x}), we turn the
Eq.(\ref{tspoint sinus}) into an integral over frequencies of
solicitation:

\begin{equation}
T \dot{S} \cong -K_I^2 \, V \int \frac{
\mu''(\omega)}{\mu'(\omega)^2+\mu''(\omega)^2}
\frac{\mbox{d}\omega}{\omega}. \label{tspoint omega}
\end{equation}

The viscous dissipation $T \dot{S}$ and $G_v$ are simply related
by $T \dot{S}=V G_v$. Finally, as the fracture energy at zero
velocity $G_0$ and the applied stress intensity factor $K_I$ are
related by the classical expression $G_0 = K_I^2 / \mu_{\infty}$,
we end up with \cite{PGG2}:

\begin{equation}
\frac{G_v(V)}{G_0} \cong \mu_{\infty}
\int_{\omega_{min}}^{\omega_{max}}
\frac{\mu''(\omega)}{\mu'(\omega)^2+\mu''(\omega)^2}
\frac{\mbox{d}\omega}{\omega}. \label{int expr}
\end{equation}
$G_0$ is the adhesion energy due to local processes near the tip.
The limiting values $\omega_{min}=V/L$ and $\omega_{max}=V/l$
define the range of frequency over which the material is excited.

The analytical results derived from Eq.(\ref{int expr}) are
presented in the appendix. Using the form (\ref{rheolaw}) of the
complex modulus of the material, the evolution of the total
adhesion energy $G= G_0 + G_v$ versus crack velocity can be
predicted for both cases of crosslinked and uncrosslinked ($\mu_0
= 0$) polymers. For the sake of conciseness, we will only
qualitatively discuss the contribution of the different zones for
$G_v$, which contrasts with some predictions of the classical
picture of viscoelastic trumpet.

\subsubsection{Contribution of the different zones for the adhesion
energy}

It was pointed out in a preceding paper \cite{PGG2} that the
dissipation in the liquid zone predominates for the overall
dissipation (because the liquid region is huge, ranging from
$V\tau$ to $\lambda V \tau$). This gives rise to amplification of
the adhesion energy by a factor $\lambda$ ($G=\lambda G_0$) when
the separation rate is taken between $l/\tau$ and $L/\lambda
\tau$. This domain of velocities corresponds to a fully developed
liquid zone, which explains the maximum of fracture energy (as the
other "elastic" zones do not dissipate energy in this simple
view).

In fact, it is worth noting that this enhancement begins for {\it
much lower separation rates}. In order to get a better
understanding for the origin of this property, let us now
calculate the viscous dissipation in the various regions of the
bulk polymer, by estimating in each zone the value of the complex
modulus: $\underline{\mu}(\omega)=\mu'(\omega)+i\mu''(\omega)$
(the different approximations are mentioned in Fig. 1).
The crucial point here is that the integration of a {\it weak}
loss modulus $\mu''$ as in Eq.(\ref{int expr}) over a {\it huge}
volume can give rise to an energy enhancement comparable to the
viscous region one. In Fig. 3, we present a simplified
graphic representation of the regimes.
\begin{figure}
\centering\resizebox{0.6 \textwidth}{!}{%
  \includegraphics*[1.5cm,10.2cm][20cm,23cm]{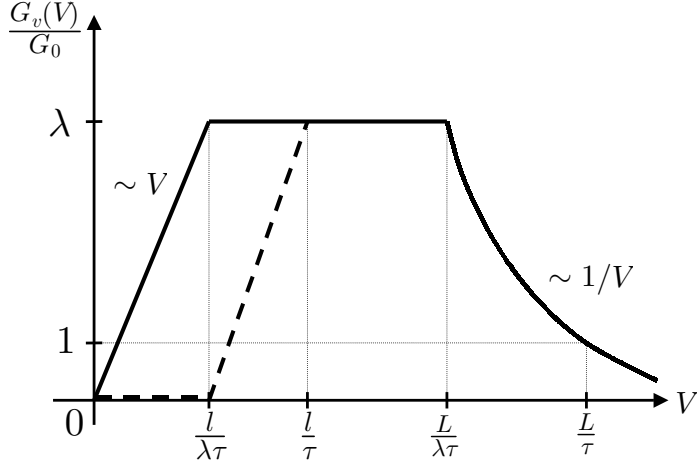} }
  \caption{Qualitative diagram of the viscous contribution $G_v$
  for fracture energy of a crosslinked polymer, versus
  separation rate $V$. The speed regime corresponding to
  Fig.2 is $l/\tau < V < L/(\lambda \tau)$
  (plateau region with an enhancement $\lambda$ of the adhesion
  energy). The limiting case $\lambda \rightarrow \infty$ of
  uncrosslinked polymers is characterized by the decreasing
  portion of curves ($G_v \sim 1/V$).
  For comparison, the dotted line represents the evolution of fracture energy as expected by
  earlier developpements \cite{PGG1,PGG2}~: it is clear that the viscous amplification of adhesion
  begins at low separation rates, even in the absence of a real "liquid" zone ($V<l/ \lambda \tau$).}
  \label{gv}
\end{figure}

At {\it low separation rates}, $V<l/(\lambda \tau)$, the whole
polymer material behaves as a "soft solid"; replacing $\mu'$ and
$\mu''(\omega)$ by their approximated expressions in Eq.(\ref{int
expr}), we obtain:

\begin{equation}
\frac{G_v(V)}{G_0} \sim \frac{\lambda^2 V \tau}{l} \; \;\mbox{for}
\; \; V<\frac{l}{\lambda \tau}. \label{low speed}
\end{equation}

Thus, the (weak) loss modulus of the elastic region is sufficient
to increase the fracture energy. The fracture energy reaches $G
\sim \lambda G_0$ for a fracture speed as low as $V=l/(\lambda
\tau)$, even if the "liquid zone" has not emerged yet.

For {\it intermediate separation rates} [$l/(\lambda
\tau)<V<L/(\lambda \tau)$], the "soft solid" region gives a
dissipation of the same order as the liquid zone:

\begin{equation}
\frac{G_v(V)}{G_0} \sim \lambda \;  \;\mbox{for} \;    \;
\frac{l}{\lambda \tau}<V<\frac{L}{\lambda \tau}. \label{interm
speed}
\end{equation}

Finally, at {\it high separation rates} [for $V>L/(\lambda
\tau)$], $G(V)$ is a decreasing function:

\begin{equation}
\frac{G_v(V)}{G_0} \sim \frac{L}{V\tau} \; \; \mbox{for} \; \; V >
\frac{L}{\lambda \tau}. \label{high speed}
\end{equation}

For uncrosslinked polymers, the rheological behavior is viscous at
low frequencies: $\mu_0=0$ and the complex modulus can be written
as:
\begin{equation}
\mu(\omega) = \mu_{\infty} \frac{i \omega \tau}{1+ i \omega \tau}.
\label{rheolaw uncross}
\end{equation}

Equation (\ref{form PGG}) thus gives:

\begin{equation}
\frac{G_v(V)}{G_0} \sim \frac{L}{V\tau}. \label{gv for uncross}
\end{equation}

As expected, we recover the expression (\ref{high speed}) for a
crosslinked polymer at velocities $V$ larger than $L/\lambda
\tau$, {\it i.e.} when the soft solid region has disappeared
because of the finite dimensions of the sample.

The adhesion energy expression (\ref{gv for uncross}) is compared
with experiments in part \ref{gv exp uncross}.

\subsection{The trumpet profile}
\label{tromp prof}

If we assume the relation (\ref{relation omega x}) between the
distance to fracture tip and the frequency imponed to the
material, we can relate the rheological properties of the material
and the corresponding profile $u(x)$ in each region. In a linear
approach of stress strain relations in the viscoelastic medium
\cite{christensen_book}, the stress $\sigma(t)$ and strain
$\gamma(t)$ are related by the relaxation modulus $G(t)$:

\begin{equation}
\sigma(t)=\int_{-\infty}^{t} G(t-\tau)
\frac{\mbox{d}\gamma(\tau)}{\mbox{d} \tau} \mbox{d}\tau.
\label{general}
\end{equation}

If the strain history is specified as being a harmonic function of
time according to $\underline{\gamma}(t)= \gamma_0 \exp{i \omega
t}$ (with an amplitude $\gamma_0$), we can write:
\begin{equation}
\underline{\sigma}(t)= \underline{\mu}(\omega) \gamma_0
\exp^{i\omega t}= |\underline{\mu}(\omega)| \gamma_0 \exp^{i(
\omega t + \tan^{-1}[\mu''(\omega)/\mu'(\omega)])},
\label{harmonic}
\end{equation}

where $|\underline{\mu}(\omega)|$ is the magnitude of the complex
modulus $\underline{\mu}(\omega)$. Taking the modulus of both
sides of Eq.(\ref{harmonic}), we get:

\begin{equation}
|\underline{\sigma}(t)|= |\underline{\mu}(\omega)| |\gamma_0|.
\label{magnitude}
\end{equation}

For a steadily growing mode I interface plane stress (or plane
strain) crack under small scale yielding conditions, all field
quantities are time independent with respect to an observer
attached to the tip of the cohesive zone. Defining a polar
coordinate system $(r,\theta)$ (cf. Fig. 2), we know
\cite{rice} that the amplitude of stress inside an elastic
material is given by:

\begin{equation}
\sigma_{ij}(r,\theta)=\frac{K_I}{\sqrt{2 \pi r}} g_{ij}(\theta),
\label{stress general}
\end{equation}

where $K_I$ is the applied stress intensity factor \cite{tada} and
$g_{ij}(\theta)$ are universal functions describing the angular
variation of the crack tip stress field. In particular, the normal
stress vanishes on the tip, but the other stress components follow
the following scaling law with respect to the distance $x$ to the
crack tip:

\begin{equation}
\sigma(x)=\frac{K_I}{\sqrt{x}}. \label{k1}
\end{equation}

Although we are dealing here with a complex viscoelastic medium,
this simple scaling form (Eq.\ref{k1}) remains valid for our
problem, as the equations of motion reduce in both cases to
$\nabla \sigma=0$, with identical compatibility conditions imposed
to the stress components \cite{PGG_book}. The strain imposed to a
fluid element of length $\mbox{d}x$ is simply given by $\gamma =
\mbox{d}u/\mbox{d}x$. Relating $x$ to $\omega$ by the scaling
expression (\ref{relation omega x}), we finally obtain:

\begin{equation}
\left| \mu \left( \omega=2 \pi \frac{V}{x} \right) \right| \cdot
\frac{\mbox{d}u}{\mbox{d}x}= \frac{K_I}{\sqrt{x}}. \label{form
PGG}
\end{equation}

In each solid or liquid zone, we know from Eqs.(\ref{muprime et
musec}) the expression of $|\mu|=\sqrt{\mu^{'2}+\mu^{''2}}$, and
can derive from Eq.(\ref{form PGG}) the expected profile $u(x)$
\cite{PGG1}.

In the {\it soft solid region}, where the frequencies are low
($\omega < 1/(\lambda \tau)$), we recover an elastic modulus:
$|\mu|= \mu_0 \sqrt{1+(\lambda \omega \tau)^2} \approx \mu_0$.
Equation (\ref{form PGG}) can be rewritten as: $\mu_0
\mbox{d}u=\sigma(x) \mbox{d}x$, which simply traduces the
linearity of displacement $\mbox{d}u$ when a force $\sigma(x)
\mbox{d}x$ is exerted on the element $\mbox{d}x$: the behavior is
elastic, and we obtain from Eq.(\ref{form PGG}) the expected
profile in region 1:

\begin{equation}
u_{\mbox{\tiny soft solid}} \sim x^{1/2}. \label{soft prof}
\end{equation}

In the {\it liquid region}, in the same manner, we recover an
viscous-type rheology, as $|\mu|= \mu_{\infty} \omega \tau
\sqrt{1+(\omega \tau)^2} \approx \mu_{\infty} \omega \tau $.
Eq.(\ref{form PGG}) can be rewritten as: $\sigma =\eta
\dot{\gamma}$, which is simply the stress constitutive relation
for a Newtonian viscous fluid. Knowing that $|\mu|\sim 1/x$ in
this zone, Eq.(\ref{form PGG}) gives the expected profile in
region 2:

\begin{equation}
u_{\mbox{\tiny liquid}} \sim x^{3/2}. \label{liquid prof}
\end{equation}

The conclusions for the {\it strong solid region} are identical to
the soft one. We predict:
\begin{equation}
u_{\mbox{\tiny strong solid}} \sim x^{1/2}. \label{strong prof}
\end{equation}

Greenwood and Johnson \cite{greenwood} investigated the precise
shape of the free surfaces of a crack, by using Barenblatt's
approach \cite{barenblatt}. Their conclusions agree with our
qualitative picture of viscoelastic fracture: at very low speed,
the crack has an elastic shape $u \sim x^{1/2}$ (\emph{i.e.} the
liquid zone has not emerged yet); at high speed, the profile is
again purely elastic (\emph{i.e.} the bulk of the material is only
a strong solid zone); finally, at intermediate speeds, the three
zones are present, including a central $u \sim x^{3/2}$ profile,
whose extension depends on the form of the compliance function for
the material.

These different velocity regimes are indeed experimentally
observed (cf. Fig. 7), as developed below. Of course, the
purely elastic shape at low velocity is not observed, as there is
no elastic modulus at low strain-rates for an uncrosslinked
polymer.

\section{Experimental results}
\label{experimental}

\subsection{Experimental setup}

In the experimental study, we chose to use an uncrosslinked
polymer instead of a poorly crosslinked elastomer. The advantage
is that such a liquid adhesive gives large enough deformations to
be evidenced by simple optics set-up. Even if all the regimes
discussed in the previous part will not be present in the
experiments, the influence of viscous flow during failure was
investigated in detail with in particular, the study of fracture
profiles. This last point also required the use of thick adhesive
layers in order to avoid additional cut-off length related to the
thickness. The results presented below correspond to an adhesive
thickness of 10 mm, larger than the diameter of the probe-adhesive
contact area (about 6 mm). The adhesive we used is an
uncrosslinked polydimethylsiloxane (PDMS) polymer (Rhodia, France)
with a large molecular weight (M$_w$ = 497000, M$_w$/M$_n$ = 1,9).
It exhibits model rheological properties and has a large
characteristic time ($\tau =$ 0.6 s) well suited to video
recording rate.

The experimental procedure is detailed in \cite{Ondarcuhu} : we
built a probe tack test experiment and adapted an optical set-up
in order to visualize and measure the fracture profiles during the
debonding process. This allowed to correlate the energy and stress
measurements with the fracture propagation mechanism. An important
point of our set-up is that we used a spherical probe (a glass
watch). This was necessary to overcome the spurious problems of
parallelism met with a flat probe. With the spherical one, the
fracture always propagated symmetrically. This was nearly never
the case with a flat probe. Note that a very small curvature is
sufficient to guide the fracture radially : we used probes with a
radius of curvature (about 10 cm) much larger than the observed
deformations (about 1 mm) and than the radius of the
probe-adhesive contact area (about 3 mm). As the characteristic
angle of aperture of the probe (of order $3/100$) is very small
compared with the opening angle of fracture (of order $1/3$), we
therefore assumed that it did not modify the fracture mechanism
compared with a flat probe. This spherical geometry is also very
convenient to lighten and get sharp images. Another feature to
note is that the velocity of the fracture measured on the video
was constant during the propagation. The propagation speed
increased only at the very end of the debonding. This is an
important point in order to compare the results to the models
developed at constant velocity. For every experiment we monitored
the force F(t) during the debonding and deduced by numerical
integration of this curve the adhesion energy per unit surface G.
We also videotaped the fracture propagation. We measured the
fracture profile $u(x)$ as the distance between the polymer
surface and the probe.

\subsection{Adhesion energy $G$ versus crack velocity}
\label{gv exp uncross} In the following we will consider only the
case of adhesive failure when the bond breaks at the polymer-probe
interface. At smaller separation velocities, the failure was
cohesive as discussed in \cite{Ondarcuhu}. In Fig.4 we
reported the adhesion energy of the bond as a function of the
velocity $V$ of propagation of the fracture. Note that the
velocity $V$ is different from the velocity at which the probe is
withdrawn. Two regimes are clearly evidenced:
\begin{figure}
\centering\resizebox{0.6 \textwidth}{!}{%
  \includegraphics*[1.6cm,12.5cm][18.8cm,24.2cm]{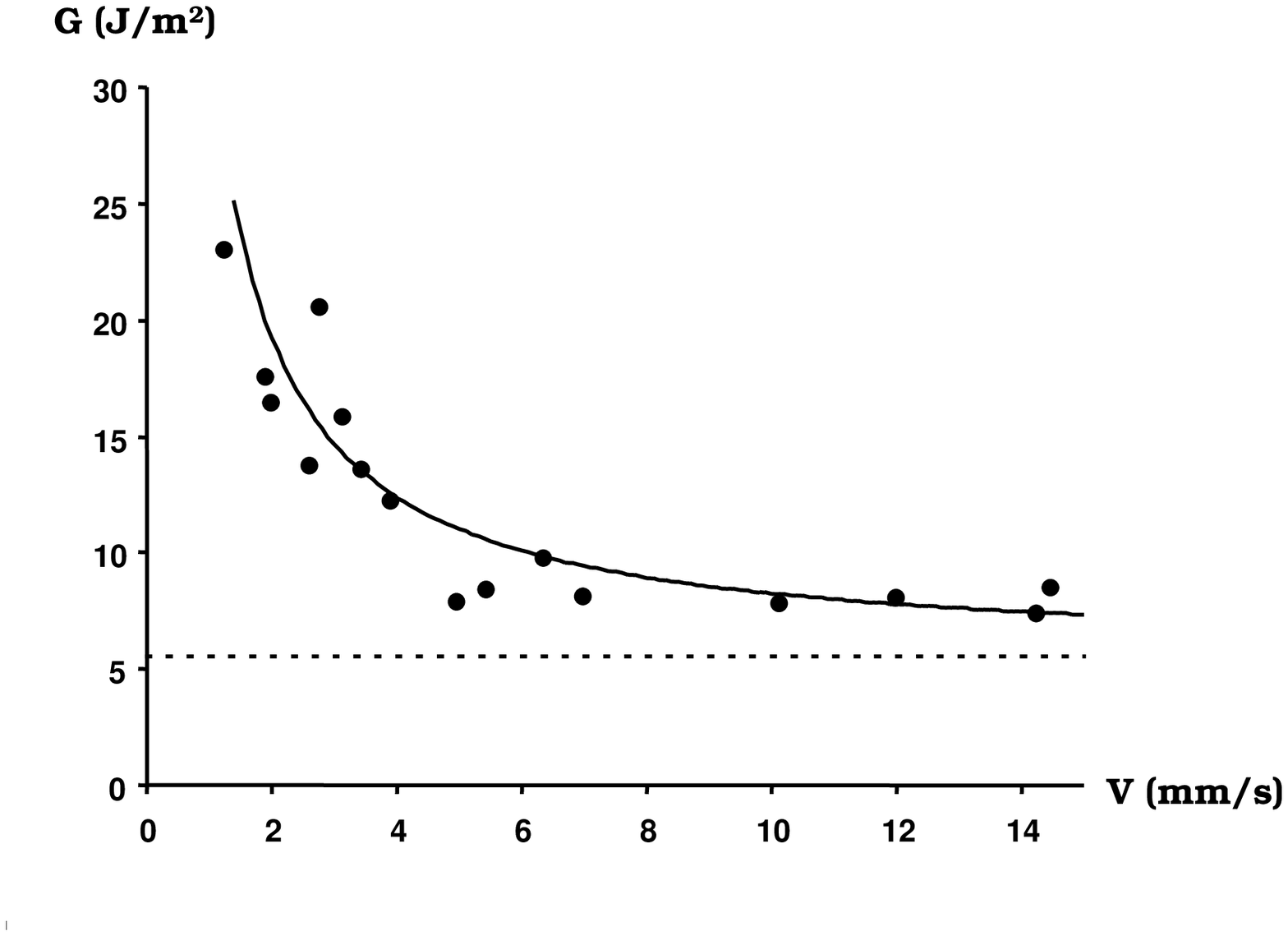} }
  \caption{Adhesion energy $G$ versus the fracture velocity $V$
  for a 10 mm-thick adhesive bond prepared during a contact time
  $T_c =$ 1s and a contact pressure $P_c =$ 0,05 N.m$^{-2}$. The
  dotted line represents the asymptotic value ($G=G_0$) reached by the
  fracture toughness at high separation rates. The solid curve corresponds
  to equation \ref{gv uncross corr}
  with $L = 3$ mm, $\tau =$ 0.6 s, and $G_0 =$ 5.5 J.m$^{-2}$.}
  \label{gvexp}
\end{figure}

i) at large velocities, the energy $G$ did not depend on the
separation rate and tends towards a constant value. The fracture
propagating at the adhesive-probe interface was very rapid so that
the polymer has no time to flow: it has an elastic behavior. In
this case the constant energy value is governed by processes
occurring close to the fracture tip.

ii) for smaller velocities the failure was still interfacial but
the energies measured are larger. As shown in \cite{Ondarcuhu}
this energy enhancement is caused by viscous losses in the bulk of
the polymer far from the fracture tip. The fracture is slow enough
to let the polymer flow.

We compared the energy measurements reported in Fig.4
with the model described above. From Eq.(\ref{gv for uncross}) we
deduced the expression of the fracture energy :
\begin{equation}
G= G_0 \left( 1+ \frac{L}{V \tau} \right). \label{gv uncross corr}
\end{equation}

As shown in Fig.4 this expression describes very well the energy
measurements. The energy enhancement appeared for fracture
velocities lower of 5 mm$\cdot$s$^{-1}$ which is in very good
agreement with the expected value $V = L/\tau$. In our experiment
$L \sim$ 3 mm, $\tau \sim$ 0.6 s which gives $V \sim$ 5
mm$\cdot$s$^{-1}$. Using these values we deduced from the fit the
value of $G_0 =$ 5.5 J$\cdot$m$^{-2}$.

\subsection{Fracture profiles}
\label{frac prof exp}

The fracture profiles measured in the moving frame during an
experiment at V = 2 mm$\cdot$s$^{-1}$ corresponding to the
viscoelastic regime are represented in Fig.5 \ . At the
beginning of the propagation the profile was parabolic $u \sim
x^{1/2}$ as evidenced in the logarithmic representation
(Fig.6). As the propagation proceeded we noted an
important modification of the profile : close to the fracture tip
the profile is still parabolic but we observe a change in the
concavity of the profile far from the fracture tip. For $x$ larger
than about 1 mm the profile is described by a power law larger
than 1/2. This distance is in good agreement with the $L \sim
V\tau =$ 1.2 mm distance expected in the theoretical part as the
transition between solid and liquid zones. All these profiles
coming from a single experiment corresponds qualitatively to the
trumpet profile discussed in detail above.
\begin{figure}
\centering\resizebox{0.6 \textwidth}{!}{%
  \includegraphics*[2.8cm,9.5cm][17cm,20.5cm]{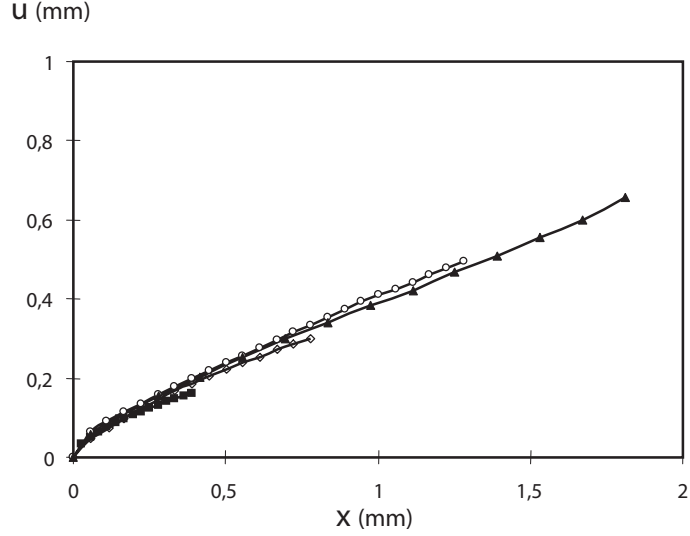} }
  \caption{Fracture profiles $u(x)$ in the moving frame at
  different times ($\blacksquare$ $t=0.2$s, $\diamond$
  $t=0.4$s, $\circ$ $t=0.6$s, $\blacktriangle$ $t=0.9$s;
  $t=$0 corresponds to the beginning of crack propagation)
  during one debonding experiment ($V =$ 2 mm$\cdot$s$^{-1}$,
  $G= 3 \, G_0$)}
  \label{ux1}
\end{figure}

\begin{figure}
\centering\resizebox{0.6 \textwidth}{!}{%
  \includegraphics*[2.8cm,10cm][17.5cm,20.5cm]{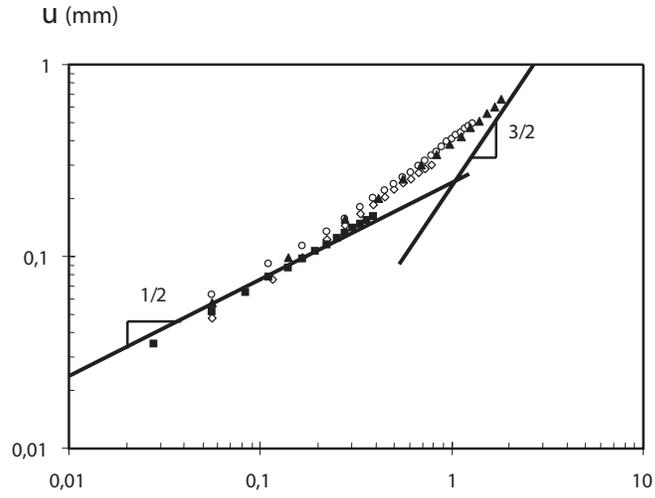} }
  \caption{Fracture profiles $u(x)$ in the moving frame at
  different times (cf. Fig.\ref{ux1}) during one debonding experiment
  ($V =$ 2 mm$\cdot$s$^{-1}$,
  $G= 3 \, G_0$), in logarithmic scale. The two black lines
  represent the power laws $u \sim x^{1/2}$ and $u \sim x^{3/2}$.}
  \label{ux1bis}
\end{figure}

It was then interesting to compare profiles of experiments coming
from the two distinct regimes : elastic and viscoelastic. In
Fig.7, we superposed two profiles obtained at about 2/3 of the
fracture propagation for an experiment in the elastic regime ($V
=$ 14.5 mm$\cdot$s$^{-1}$, $G = 1.3 G_0$) and an experiment in the
viscoelastic regime ($V =$ 3.9 mm$\cdot$s$^{-1}$, $G = 2.2 \,
G_0$). This latter profile is similar to the ones reported in
Fig.5 except that the transition from elastic to viscous profile
occurs for a larger distance $x \sim 2$ mm which is still in good
agreement with the $L \sim V \tau =$ 2.3 mm value. As for the
profile in the elastic regime, the profiles are identical close to
the fracture tip but differ in the far field region. In the
elastic case the fracture propagation is so rapid that the polymer
remains elastic. The value of $L \sim V \tau =$ 8.7 mm is larger
than the contact size. During all the propagation the fracture
profile is described by the parabolic profile $u \sim x^{1/2}$
(see Fig.8). The difference between the two curves of Fig.7
materializes the viscous flow which is responsible for the energy
enhancement obtained in the viscoelastic regime.
\begin{figure}
\centering\resizebox{0.6 \textwidth}{!}{%
  \includegraphics*[3cm,9cm][17.5cm,20.5cm]{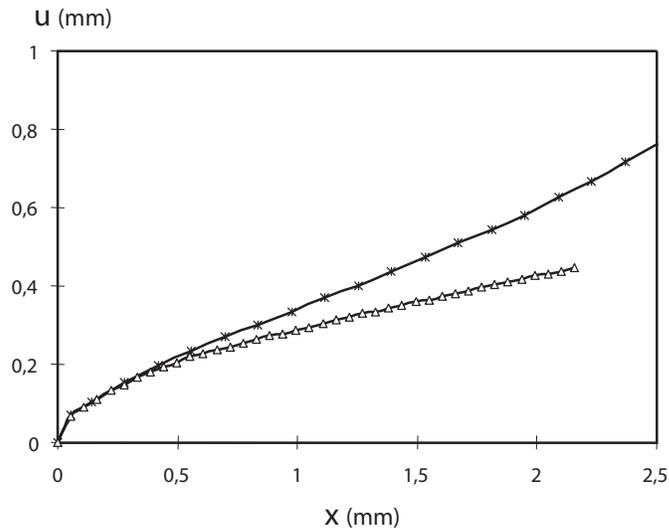} }
  \caption{Fracture profiles $u(x)$ in the moving frame corresponding
  to different experiments : ($\divideontimes$) experiment in the viscoelastic
  regime $V =$ 3.9 mm$\cdot$s$^{-1}$, $G= 2.2 \, G_0$;
  ($\vartriangle$) experiment in the elastic regime
  $V=$ 14.5 mm$\cdot$s$^{-1}$, $G= 1.3 \, G_0$.}
  \label{ux2}
\end{figure}

\begin{figure}
\centering\resizebox{0.6 \textwidth}{!}{%
  \includegraphics*[3cm,9.5cm][17.6cm,20.5cm]{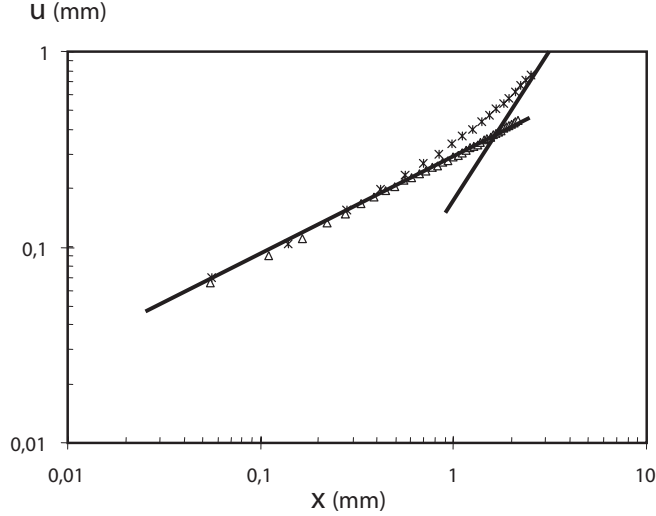} }
  \caption{Fracture profiles $u(x)$ in the moving frame corresponding
  to different experiments, in logarithmic scale: ($\divideontimes$) experiment in the viscoelastic
  regime $V =$ 3.9 mm$\cdot$s$^{-1}$, $G= 2.2 \, G_0$;
  ($\vartriangle$) experiment in the elastic regime
  $V=$ 14.5 mm$\cdot$s$^{-1}$, $G= 1.3 \, G_0$. As in Fig.\ref{ux1bis}, the black lines
  represent the power laws $u \sim x^{1/2}$ and $u \sim x^{3/2}$.}
  \label{ux2bis}
\end{figure}

All these results verify qualitatively the picture of the trumpet
profile schematized in Fig.2 (except for the soft solid
zone which is not present with a liquid polymer). However, the
range of distances available experimentally was not sufficient to
clearly observe the $u \sim x^{3/2}$ characteristic of the liquid
zone. We only observed a transition regime that can be large
because, in such systems, the transition between solid and liquid
behavior extends over several orders of magnitude of solicitation
rate.

In order to make more quantitative comparison we tried to fit the
fracture profiles with Eq.(\ref{form PGG}) which gives a direct
relation between the fracture profile $u(x)$ and the rheological
property $|\mu(\omega)|$ of the adhesive. The rheological data
$\mu'(\omega)$ and $\mu''(\omega)$ were measured using a
Rheometrix RDA2 machine. For each point of the profile, we
calculated the corresponding pulsation $\omega=2 \pi V/x$ where
$V$ is the fracture velocity and determined the corresponding
modulus $|\mu(\omega)|$. The profiles $u(x)$ were determined as
explained above. We tested two methods to derivate the profiles: a
numerical derivation or a fit by a polynomial function followed by
an analytical derivation. These two methods gave the same results
within a 10\% uncertainty. We preferred the numerical method
because the polynomial fit was not perfect close to the fracture
tip.

In Fig.9 we reported, for every profile of
Figs.5 and 7 the value of $|\mu(\omega)| \cdot
du/dx$ as a function of $x$. We observe that the curves
corresponding to the different profiles fit on a same curve. The
parabolic profiles of Fig.5 observed at the beginning of
the propagation as well as the "trumpet" profiles that develop
after a given time give the same power law in this representation.
It is also the case for the profiles of Fig.7 coming from
different types of fracture propagation. In fact, Eq.(\ref{form
PGG}) holds for all conditions of solicitation. The stress
intensity factor $K_I$ characterizes the strength of the bond and
is directly related to $G_0$ by:

\begin{equation}
G_0= \frac{K_I^2}{\mu_{\infty}}. \label{thierry}
\end{equation}

\begin{figure}
\centering\resizebox{0.65 \textwidth}{!}{%
  \includegraphics*[1.3cm,10.7cm][17cm,20.7cm]{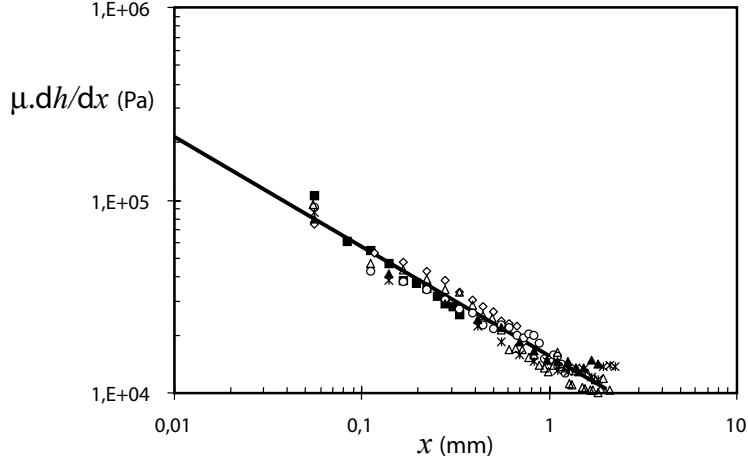} }
  \caption{$|\mu(\omega)| \cdot du/dx$ versus $x$ for the six profiles reported
  in Figs.5 and 7. The symbols used are identical to
  those used in Figs.5 and 7. The black line is the best
  power-law fit: $|\mu(\omega)| \cdot du/dx \sim x^{-0.56}$.}
  \label{mudhdx}
\end{figure}

Both parameters $G_0$ and $K_I$ describe processes occurring at
the fracture tip and therefore depend only on the preparation of
the bond. All the experiments of Fig.4 thus correspond
to a same value of $K_I$. The far field losses induced an
enhancement of the energy in the viscoelastic regime but did not
modify the stress distribution in the adhesive. For that reason,
the profiles of Figs.5-8, in the elastic as
well as in the viscoelastic regimes, coincide in the
representation $|\mu(\omega)| \cdot du/dx$ versus $x$. The power
law deduced from Fig.9 is $|\mu(\omega)| \cdot du/dx
\sim x^{-0.56}$ in rather good agreement with the $x^{-1/2}$
expected from Eq.(\ref{form PGG}). The deduced value of the stress
intensity factor is $K_I =$ 18000 N$\cdot$m$^{-3/2}$. The value
$G_0 =$ 3 J$\cdot$m$^{-2}$ deduced from the profiles according to
Eq.(\ref{thierry}) is of the order of the value $G_0 =$ 5.5
J$\cdot$m$^{-2}$ deduced from adhesion energy measurements.

All these experimental results validate the picture of the trumpet
profile and its influence on the adhesion energy of a viscoelastic
polymer.

\section{Conclusion and perspectives}

In conclusion, we proposed a qualitative description of the
dissipative processes occurring during the failure of the
interface between a viscoelastic material and a solid substrate,
mainly based on de Gennes' initial approach \cite{PGG2}. In
particular, we analyzed the enhancement of fracture energy due to
far-field viscous dissipation in the bulk material. The results of
our scaling analysis are in accordance with the more rigorous ones
of Hui \emph{et al.} \cite{hui_xu_kramer} and Greenwood \emph{et
al.} \cite{greenwood}.

We have shown that, for a crosslinked elastomer, the interface
toughness $G(V)$ starts from a relatively low value $G_0$ due to
local dissipative processes near the crack tip, and reaches a
maximum of order $G_0 (\mu_{\infty}/\mu_0)$ (see
eq.\ref{rheolaw}). It was shown that our simple model accounts
well for the dependence of fracture energy $G(V)$ for a polymer
melt, for which the fracture energy should scale as $1/V$. The two
velocity regimes, and the peculiar crack profiles experimentally
observed validate our description in two zones: an elastic region
near the crack tip, and a liquid zone with a different concavity
at distances larger than $V \tau$ (where the material had enough
time to flow).

The "viscoelastic trumpet" model \cite{PGG2} is thus a simple,
tractable, linear model which describes well, on the basis of
viscoelasticity, the viscous origin of fracture energy enhancement
observed in many systems. Our aim here is to point out that the
fracture energy originating from viscous dissipation in the bulk
polymer had been underestimated at low velocities. Some
experiments measuring the quantity $G_0$ for elastomers, by taking
a crack velocity as small as practically possible, which obtain
too large values for $G_0$, may thus be reassessed by taking into
account the viscous dissipation existing in large volumes of soft
elastic regions.

To conclude, let us insist on the fact that this "viscoelastic
trumpet" model is a qualitative approach of a very complex
phenomenon, and lies on many approximations: in particular, the
independence versus $V$ of the adhesive zone scale $l$, and of the
associated dissipation term $G_0$, is obviously discutable as it
has been proved that for many systems $G_0$ exhibits a marked
dependence on $V$, and most of the rate dependence of $G(V)$ then
originates from the rate dependence of $G_0$ itself \cite{gent}.
We can thus expect our model to apply to systems characterized by
a relatively weak adhesion, for which $G_0$ (and subsequently $l$)
is effectively quasi-independent on $V$: this is the case of the
very high molecular weight PDMS studied in the experimental
section (at short times, or high separation rates, this type of
material behaves elastically and does not dissipate much near the
crack tip, as a consequence of its very low glass transition
temperature). Most of other uncrosslinked systems are
characterized by an increasing dissipation with increasing
velocity, even in the elastic regime \cite{benyahia_verdier,derail_allal,
christensen_everland,christensen_flint}, because the
increase in $G_0$ dominates when the adhesion hysteresis is large
and the material is soft, even when the polymer is macroscopically
elastic \cite{maugis_barquins}. A more realistic description of
the microscopic processes taking place in the adhesive zone,
including nonlinear effects and specific properties like crack
blunting \cite{HuiJagota} for very soft materials (characterized
by a weak elastic modulus $E \lesssim 0.1$ MPa), represents a
theoretical challenge, and a possible further extension of our
simplistic model.

\section{Acknowledgements}
We are very grateful to P.-G.~de Gennes, C.~Creton for stimulating
and helpful discussions. We also thank L.~L\'{e}ger, K.~Okumura,
M.~Portigliatti and C.~Poulard for useful comments.

\appendix

\section{Analytical results for crosslinked and uncrosslinked polymers}

This appendix provides the main analytical results about the
fracture energy derived from the integral expression (\ref{int
expr}) of $G(V)$. We will discuss both cases of crosslinked and
uncrosslinked polymers, whose corresponding mechanical models are
represented in Fig.10.
\begin{figure}
\centering\resizebox{0.48 \textwidth}{!}{%
  \includegraphics*[2.7cm,8.5cm][18.5cm,20cm]{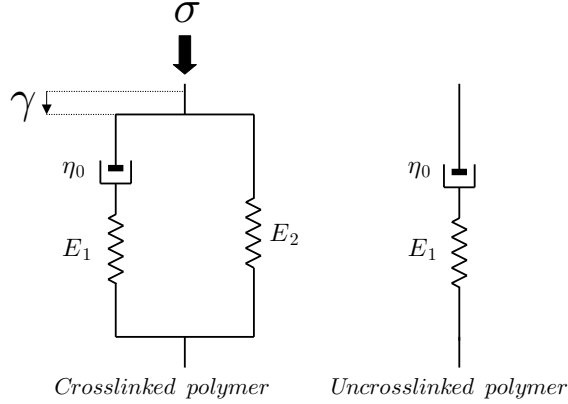} }
  \caption{ Rheological models for (i) crosslinked and (ii) uncrosslinked polymers.
  \textbf{(i)} The mechanical analog for a crosslinked material (Zener's model) consists in the association of two
  springs (with moduli $E_1$ and $E_2$), with a dashpot of viscosity $\eta$. \textbf{(ii)}
  When the second spring
  (of modulus $E_2$) is removed, we find back a liquid-like material
  (Maxwell's model), which can figure the rheology
  of a uncrosslinked polymer melt.}
  \label{modeles rheologiques}
\end{figure}

\subsection{Rheological characteristics}
To begin with, let us find back the expressions (\ref{rheolaw})
and (\ref{rheolaw uncross}) of the complex modulus for both cases.
Following the notations $E_1$, $E_2$ and $\eta$ introduced in
Fig.10, it can easily be shown that the
applied stress $\sigma$ and the strain $\gamma$ are related by the
following differential equation:

\begin{equation}
\sigma + \frac{\eta_0}{E_1} \dot{\sigma} = E_2 \varepsilon +
\eta_0 \frac{E_1 + E_2}{E_1} \dot{\varepsilon}
\label{eq:EquationDifferentielleZener}
\end{equation}

The prefactor $\eta_0 / E_1$ is a characteristic time of the
material, denoted $\tau = \eta_0 / E_1$.

\begin{equation} \underline{\sigma} = \left( E_2 + E_1 \frac{i \omega
\tau}{1 + i \omega \tau} \right) \underline{\gamma} \end{equation}

Defining the two parameters $\mu_0$ and $\mu_{\infty}$ as $E_1
\equiv \mu_{\infty}- \mu_0$ and $E_2 \equiv \mu_0$, we thus obtain
the expression (\ref{rheolaw}) of the complex modulus:

\begin{equation}
\underline{\mu}(\omega) = \mu_{0}+ (\mu_{\infty}-\mu_0) \frac{i
\omega \tau}{1+ i \omega \tau}.
\end{equation}

In the same way, the rheological behavior of an uncrosslinked
polymer, whose corresponding mechanical model is represented in
Fig.10, can be described by the
following complex modulus:

\begin{equation}
\mu(\omega) = \mu_{\infty} \frac{i \omega \tau}{1+ i \omega \tau}.
\end{equation}

\subsection{Fracture energy $G(V)$ derived from the integral
expression Eq.\ref{int expr}}

\subsubsection{Analytical results for an elastomer}

It is straightforward to see that for the rheological law
(\ref{rheolaw}), one has:

\begin{equation}
\left\{ \begin{array}{ll} \mu'(\omega)= \mu_0
+(\mu_{\infty}-\mu_0) \frac{\omega^2 \tau^2}{1+ \omega^2 \tau^2}, \\
                          \mu''(\omega)= (\mu_{\infty}-\mu_0)
                          \frac{\omega \tau}{1+ \omega^2 \tau^2}.
                    \end{array}
            \right.
\label{muprime et musec}
\end{equation}

According to Eq.\ref{int expr}, this leads to:

\begin{equation}
\frac{G_v(V)}{G_0} \cong \frac{\mu_{\infty}
(\mu_{\infty}-\mu_0)}{\mu_0^2} \tau \int_{V/L}^{V/l}
\frac{\mbox{d}\omega}{1+\lambda^2 \omega^2 \tau^2}, \label{int
expr zener}
\end{equation}

that is:
\begin{equation}
\frac{G_v(V)}{G_0} \cong (\lambda -1)
\left[\arctan\left(\frac{\lambda V
\tau}{l}\right)-\arctan\left(\frac{\lambda V
\tau}{L}\right)\right]. \label{arctan form}
\end{equation}

As pointed out before, the viscoelastic effect gives a
multiplicative enhancement of the energy $G_0$ dissipated by local
processes in the adhesive zone. The multiplicative factor, of
order $\lambda=\mu_{\infty}/\mu_0$, depends on the degree of
cross-linking: a higher crosslinkage induces an increase in
$\mu_0$ and therefore a decrease in energy of adhesion, as shown
by Gent and Petrich \cite{gent_petrich}. We have already mentioned
that a large ratio $\lambda$ can be experimentally achieved, to
give rise to a good viscous amplification of the adhesion energy.

In Fig.11 we present a logarithmic representation of the fracture
energy versus the dimensionless separation rate $\lambda V \tau
/l$. The fracture toughness $G(V)$ exhibits a maximum for a
separation rate $V^*$ given by:
\begin{equation}
V^*=\frac{\sqrt{l L}}{\lambda \tau}. \label{crit speed}
\end{equation}
\begin{figure}
\centering\resizebox{0.65 \textwidth}{!}{%
  \includegraphics*[1cm,9.5cm][19.7cm,20.6cm]{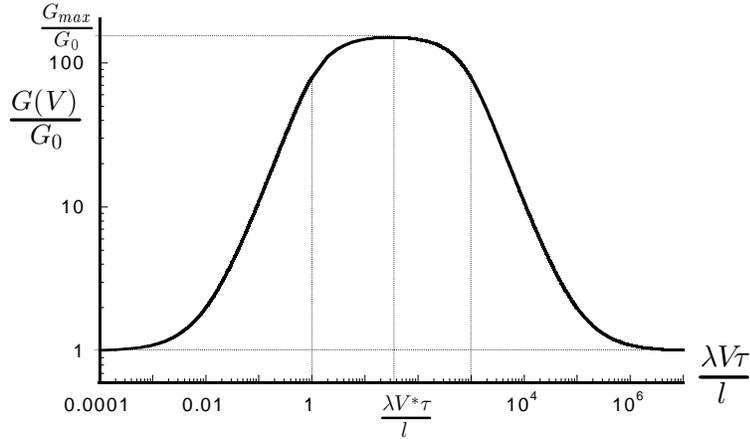} }
  \caption{Fracture energy $G(V)=G_0+G_v(V)$ versus
  separation rate $V$, according to Eq.(\ref{arctan form}) with Zener's model.
  This logarithmic representation clearly exhibits the plateau region
  with an enhancement of order $\lambda$ of the adhesion energy $G_0$. The
  curve has a maximum $G_{max}$ obtained for $V=V^*$. The ratios
  $L/l=1000$ and $\lambda=100$ were chosen for this graphic representation.}
  \label{arctan fig}
\end{figure}

The adhesion energy is maximal for $V=V^*$, where it takes the
value:
\begin{equation}
G_{max}=G_0 \left(1+(\lambda -1) \left[
\arctan\left(\sqrt{\frac{L}{l}}\right)-\arctan\left(\sqrt{\frac{l}{L}}\right)
\right] \right). \label{adh max}
\end{equation}

\subsubsection{Analytical results for a polymer melt}

For uncrosslinked polymers, characterized by the complex modulus
(\ref{rheolaw uncross}), equation (\ref{form PGG}) gives:

\begin{equation}
\frac{G_v(V)}{G_0} \sim \frac{L-l}{V \tau}. \label{gv exact for
uncross bis}
\end{equation}

Provided that the length $l$ of the adhesion zone remains small
compared with the whole sample dimension $L$, we recover the
expression \ref{high speed} for a crosslinked polymer at
velocities $V$ larger than $L/\lambda \tau$, {\it i.e.} when the
soft solid region has disappeared because of the finite dimensions
of the sample~:

\begin{equation}
\frac{G_v(V)}{G_0} \sim \frac{L}{V \tau}. \label{gv approche for
uncross bis}
\end{equation}

The adhesion energy expression (\ref{gv for uncross}) is compared
with experiments in part \ref{gv exp uncross}.


%
%


\begin{thebibliography}{9}

\bibitem{hugh_brown}
Brown, H. \emph{Physics World}  \textbf{1996 }, \emph{Jan.}, 38.

\bibitem{jones_richards}
Jones,  R. A. L. ; Richards, R. W.  \emph{Polymers at surfaces and
interfaces}; Cambridge Univ. Press, 1999.

\bibitem{creton_kramer_brown_hui}
C. Creton, C; Kramer, E. J.;  Brown, H. R.;  Hui, C.-Y.  \emph{Adv.
Polym. Sci.}  \textbf{2001},   \emph{156}, 53.

\bibitem{gent}
Gent, A. N.   \emph{Langmuir}  {\bf 1996},  \emph{12},  4492.

\bibitem{gent_shultz}
Gent A. N. ; Shultz J,  \emph{J. Adhesion}, {\bf 1972}, \emph{3}, 281.

\bibitem{andrews_kinloch}
Andrews E. H.; Kinloch A. J.  \emph{Proc. Roy. Soc. (Lond.)} {\bf 1973} \emph{A332 385}, 401.

\bibitem{wlf}
Williams, M. L. ; Landel, R. F.;  Ferry,  J. D.;  \emph{J. Am. Chem. Soc.} {\bf 1995}
\emph{77}, 3701.

\bibitem{PGG1}
de Gennes, P.-G. \emph{C. R. Acad. Sci. Paris} {\bf 1988}, {\emph 307}, 1949.

\bibitem{PGG2}
de Gennes, P.-G.  \emph{Langmuir}  {\bf 1996}, \emph{12}, 4497.

\bibitem{hui_xu_kramer}
Hui, C.-Y. ; Xu, D.-B.;  Kramer, E. J.  \emph{J. Appl. Phys.} {\bf 1992}, {\emph72},
3294.

\bibitem{christensen_and_wu}
Christensen R. M.; Wu, E. M. \emph{Engineering Fracture Mechanics}
{\bf {1981}}, \emph{14},  215.

\bibitem{bowen_knauss}
Bowen J.M.; Knauss, W. G.  \emph{J. Adhes.} {\bf{1992}},  {\emph 39}, 43.

\bibitem{christensen}
Christensen, R. M.  \emph{Int. J. Fracture} {\bf 1979} {\emph 15}, 3.

\bibitem{schapery}
Schapery, R. A. \emph{Int. J. Fracture} {\bf{1975}}  {\emph{11}}, 141;
Schapery, R. A. \emph{Int. J. Fracture} {\bf{1975}},  {\emph{11}}, 369;
Schapery, R.A. \emph{Int. J. Fracture},  {\emph{11}}, 549.

\bibitem{freund_hutchinson}
Freund, L. B.; Hutchinson, J. W.  \emph{J. Mech. Phys. Solids} {\bf{1985}}
\emph{33}, 169.

\bibitem{barber}
Barber M.;  Donley J.;  LangerJ. S. \emph{Phys. Rev. A} {\bf{1989}}  \emph{40},
366.

\bibitem{greenwood}
Greenwood, J. A.;  Johnson K. L. \emph{Phil. Mag. A.} {\bf 1981} {\emph 43}, 697.

\bibitem{PGG_pegosite}
de Gennes, P.-G.  \emph{C. R. Acad. Sci. Paris} {\bf 1991} {\emph 312}, 1415.

\bibitem{gent_kim}
Gent A. N.;  H. J. Kim H. J.  \emph{Rubber Chem. Tech.} {\bf 1990}, {\emph 63}, 613.

\bibitem{creton_leibler}
Creton, C;  Leibler, L. \emph{J. Polym. Sci. B}  {\bf 1996} {\emph 34}, 545.

\bibitem{gay_leibler}
Gay, C.; Leibler \emph{Physics Today}, {\bf 1999}, \emph{Nov} 48 ; see also
references therein.

\bibitem{Ondarcuhu}
Ondar\c{c}uhu, T. \emph{J. Phys. II France}, {\bf 1997}, {\emph 7}, 1893.

\bibitem{relaxation_time}
Equation \ref{rheolaw} corresponds to a Zener model with a {\it
single} relaxation time $\tau$. The latter assumption is rather
crude, as the dangling ends have a wide distribution of length,
etc., but allows a clear representation of the different spatial
zones far behind the fracture tip.

\bibitem{gent_lai}
Gent A. N.; Lai, S.-M.  \emph{J. Polym. Sci. B} {\bf 1994}, {\emph 32}, 1543.

\bibitem{length_L}
More precisely, $L$ is the overall length of the solicitated
region. For instance, if the viscoelastic material is in the form
of a thin slab of thickness $w$, then $L \sim w$.

\bibitem{local}
Various phenomena may occur in the adhesive zone \cite{PGG_canadian},
\emph{e.g.} chain pull-out \cite{creton_brown_schull,
leger}. In refs. \cite{hui_xu_kramer, PGG_raphael}, the material in
the adhesive zone is described a newtonian fluid with a cut-off
stress required for the interface to initiate an opening
displacement. The precise description of the adhesive zone will
not be considered here \cite{marciano_raphael, xu_hui_cramer_creton}.

\bibitem{gent_petrich}
Gent A. N.; Petrich, R. \emph{Proc. R. Soc. London} {\bf 1969}, {\emph A31}, 433.

\bibitem{christensen_book}
Christensen R. M.  {\it Theory of viscoelasticity, an
introduction}, Academic Press, 2nd edition, 1982.

\bibitem{rice}
Rice J. R., {\it Fracture II, an Advanced Treatise},  H.
Liebowitz ed., Academic Press, New York, 1968.

\bibitem{tada}
H. Tada, P. C. Paris, and G. R. Irwin, {\it The stress analysis of
cracks Handbook} del Research, Hellertown, 1973.

\bibitem{PGG_book}
de Gennes, P.-G.  \emph{Soft Interfaces} Cambridge Univ. Press,
1997.

\bibitem{barenblatt}
Barenblatt G. I. \emph{Adv. Appl. Mech.} {\bf{1962}}  \emph{7}, 55.

\bibitem{PGG_canadian}
de Gennes, P.-G. \emph{Can. J. Phys.} {\bf 1990}, {\emph 68}, 1049.

\bibitem{creton_brown_schull}
Creton, C;  Brown, H. R.;  Schull K. R. \emph{Macromol.}  {\bf 1994} {\emph 27}, 3174.

\bibitem{leger}
L\'{e}ger, L; Rapha\"{e}l, E; Hervet, H. \emph{Adv. in Polym. Sci.} {\bf 1999}
\emph{138}, 185.

\bibitem{PGG_raphael}
Rapha\"{e}l, E; de Gennes P.-G.  \emph{J. Phys. Chem.} {\bf 1992} {\emph 96}, 4002.

\bibitem{marciano_raphael}
Marciano, Y.;  Rapha\"{e}l, E. \emph{Int. J. Fracture} {\bf 1994} , {\emph 67}, 23.

\bibitem{xu_hui_cramer_creton}
Xu, D.-B.;  Hui, C.-Y.; Kramer, E. J.; Creton, C.  \emph{Mechanics of
Materials} {\bf 1991}, {\emph {11}}, 257.

\bibitem{benyahia_verdier}
Benyahia, L.;  Verdier, C; Piau, J. M.  \emph{J. Adhes.} {\bf 1997}, {\emph 62}, 45.

\bibitem{derail_allal}
Derail, C; Allal, A.; Marin, G.; Tordjeman, P. \emph{J. Adhes.} {\bf 1997} {\emph
61}, 123.

\bibitem{christensen_everland}
Christensen, S. F. ;  Everland, H.; Hassager, O.; Almdal, K. \emph{Int.
J. Adhesion and Adhesives} {\bf 1998} {\emph 18}, 131.

\bibitem{christensen_flint}
Christensen, S. F. ;  Flint, S. C.  \emph{J. Adhes.} {\bf 2000}, {\emph 72}, 177.

\bibitem{maugis_barquins}
Maugis, D; Barquins, M. \emph{J. Phys. D: Appl. Phys.} {\bf 1978},  {\emph 11}, 1989.

\bibitem{HuiJagota}
Hui, C.-Y.; Jagota, A; Bennison, S. J.;  Londono, J. D. \emph{ Proc.
Royal Soc. London, Ser. A: Mathematical and physical sciences} {\bf 2003} {\emph
403}, 1489.

\end{thebibliography}
\end{document}